\documentstyle[preprint,aps]{revtex}
\tighten
\begin{document}
\draft
\title{Resolution of the Solar Neutrino Anomaly}
\author{Maurice Dubin$^1$ and Robert K. 
Soberman$^2$\protect\cite{*}} 
\address{$^1$14720 Silverstone Dr., Silver Spring, MD 20905\\  
$^2$2056 Appletree St., Philadelphia, PA 19103}
\date{\today}

\maketitle

\begin{abstract}
The solar neutrino anomaly, measurements discrepant from 
predictions of the Standard Solar Model, has existed for 
over 30 years.  Multiple experiments measuring fluxes from 
several reactions in the hydrogen fusion chain have added to the 
puzzle.  Each of the several elements of the enigma are resolved 
by recognition of measurements establishing that most of the 
sun's fusion must occur near the surface rather than the core. 
\end{abstract}

\narrowtext

\noindent THE ENIGMA -- Five elements presently comprise the 
solar neutrino puzzle; all consequences of divergence between 
Standard Solar Model (SSM) prediction and measured data: 1) 
neutrino detection shortfall; 2) neutrino solar cycle variation; 
3) two neutrino detectors with discrepant results; 4) neutrino 
surges coinciding with major solar flares, and 5) absence of a 
measurable $\rm ^7Be$ neutrino flux.  All present measurements 
are consistent with near surface fusion and require no new 
physics nor assumed neutrino properties. 

Initial results of the first substantial solar neutrino 
detector\cite{Davis68} showed the flux well below the SSM 
predicted value.  While later theory changes reduced the 
discrepancy from ${{1} \over {10}}$ to ${{1} \over {3}}$, the 
difference remains beyond uncertainties\cite{Bahcall89}.  In 
consequence, that radio-chemical $\rm ^{37}\hskip -1pt Cl$ 
Homestake experiment remains under scrutiny.  Others, existent 
and planned, seek to further the study\cite{Bahcall95}.  The 
Kamiokande \v Cerenkov ``telescope'' is sensitive to only higher 
energy neutrinos\cite{Hirata91}, while two radio-chemical $\rm 
^{71}Ga$ experiments, SAGE\cite{Abazov91} and 
GALLEX\cite{Anselmann92} are able to detect comparatively low 
energy neutrinos.  All have measured neutrino fluxes 
substantially below expectation.  Recently SAGE 
reported\cite{Abdurashitov94} a mean since 1990 start-up of 73 
solar neutrino units (SNU), while GALLEX reduced their mean since 
1991 start-up, 87 SNU earlier\cite{Anselmann93} to 77 
SNU\cite{Anselmann95}. Solar neutrino data are averaged as 
accepted theory forbids variance in times less than 10 million 
years\cite{Bahcall89}.  SAGE and GALLEX measurements, dominated 
by neutrinos from the fundamental proton-proton (PP) fusion 
reaction (Fig.~1), are $\sim\! 60\%$ of prediction. Both 
GALLEX\cite{Anselmann95} and SAGE\cite{Abdurashitov94} report 
the flux below the predicted\cite{Bahcall85} level of 79 SNU 
(Fig.~2) at which the sun is fusing the minimum amount of 
hydrogen required for its radiation.  Mean for both is currently 
$\rm 75 \pm 7~SNU$.  

The second element is a reported $\sim\! 11$ year variation of 
the Homestake measured neutrino flux anti-correlated with sunspot 
number\cite{Rowley85}.  Neutrino capture produces only $\sim\! 
0.5$ radioactive $\rm ^{37}\hskip -2pt Ar$ atom/day, hence 
individual measurements fluctuate ($1\,\sigma\,\approx 30\%$).  
With solar gravitational and luminosity time constants 
\hbox{$\sim\! 10^7$} and nuclear \hbox{$\sim\! 10^{10}$ years}, 
SSM prohibits short period variation,  with reported changes 
attributed to statistical scatter abetted by analysis 
procedures\cite{Bahcall87}. 

The third element is an apparent discrepancy between Homestake 
results ($\sim\! 35\%$ of SSM), believed dominated by neutrinos 
from $\rm ^8B$ (Fig.~1) and Kamiokande ($\sim\! 50\%$ of SSM) 
that should measure, almost exclusively, the same $\rm ^8B$ 
neutrinos.  Davis\cite{Davis94}, has already shown that the 
difference disappears if only contemporaneous acquired data are 
compared.  

The fourth element involves neutrino flux increases noted in 
Homestake results coinciding with major solar 
flares\cite{Baziliv82}.  These surges (SSM incompatible) are 
explained as statistical fluctuation or cosmic ray 
produced\cite{Bahcall89}.  Computer simulations show the $\rm 
^{37}\hskip -1pt Cl(\nu, e^-)^{37}\hskip -2pt Ar$ detection rate 
of $\sim\! 1.2$~atoms/day associated with flares relative to a 
mean of $\rm 0.5$~atoms/day can be expected statistically in 
several percent of the runs\cite{Davis94}. 

The correlation between a great solar flare and Homestake 
neutrino enhancement was tested in 1991. Six major flares 
occurred from May~25 to June~15 including the great June~4 flare 
associated with a coronal mass ejection and production of the 
strongest interplanetary shock wave ever recorded (later detected 
from spacecraft at 34, 35, 48, and 53~AU)\cite{Gurnett93}.  It 
also caused the largest and most persistent (several months) 
signal ever detected by terrestrial cosmic ray neutron monitors 
in 30 years of operation\cite{Webber93}.  The Homestake exposure 
(June 1--7) measured a mean $\rm ^{37}\hskip -2pt Ar$ production 
rate of $3.2 \pm 1.5$ atoms/day ($\rm \approx\! 19\ ^{37}\hskip -
2pt Ar$ atoms produced in 6 days)\cite{Davis94}; about 5 times 
the rate of $\rm \approx\! 0.65\ day^{-1}$ for the preceding and 
following runs, $>\! 6$ times the long term mean of $\rm 
\approx\! 0.5\ day^{-1}$ and $>\! 2 {{1} \over {2}}$ times the 
highest rates recorded in $\sim\! 25$ operating years.  
Attributing this burst (SSM forbidden) that dwarfed the 
background neutrino flux to statistical variation stretches 
probability when it coincides with the largest recorded flare.  

The fifth element has been termed\cite{Bahcall94} a ``paradox.''  
Kamiokande results are about 50\% of the 
predicted\cite{Bahcall92} $\rm ^8B$ neutrino flux\cite{Hirata91}.  
Applying that percentage to the SSM prediction of 7.9~SNU for 
Homestake exceeds their mean of $\rm 
2.55\,\pm\,0.25$~SNU\cite{Bahcall95} leaving no room for $\rm 
^7Be$ neutrino detections (Table~I).  Results from both 
GALLEX\cite{Anselmann95} and SAGE\cite{Abdurashitov94} below 
the predicted PP neutrino flux (almost model 
independent) also leave no room for a $\rm ^7Be$ neutrino flux.  
As $\rm ^8B$ results from $\rm ^7Be$ (Fig.~1), measuring 
neutrinos from the daughter reaction while absenting those from 
the parent is paradoxical.   

Revised ``Mikheyev-Smirnov-Wolfenstein (MSW)'' neutrino physics\cite{MSW} explains the discrepancy 
between SSM prediction and observation (Table~I) by hypothetical 
neutrino flavor changes.  Measurement has excluded almost all 
possible MSW ``solutions.''  Several ad hoc non-standard solar 
models that mix various fractions of the core continuously or 
episodically\cite{Sienk90} represent other efforts to match 
measurement. If the data are accepted, invented physics models 
cannot explain neutrino surges with flares and/or solar activity 
variation. 

The solar neutrino puzzle is the SSM conflicting with 
observation!  After summarizing a set of measurements that show 
most (all?) solar fusion occurring near the surface 
(contradicting a fundamental SSM assumption), we present our near 
surface fusion interpretation that resolves {\it all} aspects of 
the solar neutrino puzzle. 

\vskip 4pt
\noindent $\rm^7Be$\ IN EARTH ORBIT REQUIRES FUSION NEAR THE 
SUN'S SURFACE -- Two independent teams measured $\rm ^7Be$ on the 
leading surfaces of the Long Duration Exposure Facility (LDEF), 
retrieved after orbiting the Earth for 69 months\cite{Fishman91}.  
The concentration necessary to account for the measurements was 
several orders of magnitude greater in orbit than in the 
stratosphere $\rm \sim\! 300\,km$ below.  In the troposphere and 
stratosphere, radioactive $\rm ^7Be$ (53 day half-life) is 
accepted as the consequence of high energy cosmic ray spalling of 
nitrogen and oxygen mostly between 15 and 20 km\cite{Arnold55}.  
With orbit altitude production inconsequential, a fast vertical 
transport and concentration mechanism was sought.  To bolster 
that hypothesis, pieces of LDEF were examined for $\rm ^{10}Be$, 
a radioisotope with a 1.5 million year half-life, similar 
chemistry, spallation production and transport likelihood.  
Unexpected, the only $\rm ^{10}Be$ found was inherent in the 
aluminum, as about the same was found on interior and control 
surfaces and it did not exhibit the $\rm ^7Be$ 100 times leading 
to trailing surface excess\cite{Gregory93}.  This makes cosmic 
ray spallation a most improbable source. 

Fusion is the remaining method to produce $\rm ^7Be$ with the sun 
the sole source consistent with known astrophysics.  Further, the 
half-life only permits the $\rm ^7Be$ to originate near the 
surface, as SSM tables\cite{Bahcall89} show core fusion $\rm 
10^{19}$ short of providing an adequate amount for the LDEF 
measurements.  Solar luminosity requires $4[10]^{38}$ protons 
fused per second.  If PP$II$ branching is $\sim\! 10\%$, this 
yields $\rm 5[10]^{36}\,$ $\rm ^7Be$ nuclei per second.  If about 
$10^{-4}$ of this joins the solar wind, it provides $\sim\! 
5[10]^{32}$ nuclei/s.  The escaping fraction may be contested 
$10^{\pm 2}$ without altering the conclusion.  This produces a 
flux at Earth of \hbox{$\rm \sim\! 2[10]^9\,atoms\cdot m^{-
2}\cdot s^{-1}$}.  The Earth's bow shock is a formidable barrier 
to the solar wind\cite{Chappell87}, so we estimate penetration of 
\hbox{$\sim\! 10^{-3}$}.  Allowing for in-flight decay, we obtain 
an influx of \hbox{$\rm \sim\! 10^{6 \pm 3} m^{-2} \cdot s^{-1}$} 
of $\rm ^7Be$ atoms.  Rather than estimate velocity and density, 
we take this flux impinging directly on the LDEF leading surface.  
If \hbox{$\sim\! 10^{-3}$} atoms stick, then we may approximate 
the equilibrium concentration by \hbox{$\rm \sim\! 10^3\,m^{-2} 
\cdot s^{-1}$} times the 53 day half-life or \hbox{$\rm \sim\! 
5[10]^9 \,m^{-2}$} as reported\cite{Fishman91}, with our 
computation having \hbox{$\sim\! 10^{\pm 4}$} uncertainty. 

Surface fusion is no longer bizarre since the \hbox{2.2~Me\hskip 
-2ptV} gamma ray line of the P(n,$\gamma$)D reaction was 
observed\cite{Terekhov93} during the solar flare of May 24 1990.  
Why would DP fusion in flares not supply a sufficient flux of 
$\rm ^7Be$ atoms?  The flux estimated above would have to be 
reduced by $\sim\! 10^4$ for the solar deuterium/hydrogen ratio 
and an additional factor of $\ge\! 10^3$ as major flares occur 
only several times per year even near the peak of the solar 
cycle.  Hence, if only flare production is invoked for the $\rm 
^7Be$, a shortfall $\ge\! 1,000$ results relative to the LDEF 
measurement.  Another method of quantifying this flare shortfall 
is to equate our low uncertainty to the deuterium/hydrogen ratio.  
Then the ratio of flare radiance to solar luminosity is the 
amount by which flare fusion fails to account for the LDEF 
measurement.  The $\rm ^{10}Be$ experiment\cite{Gregory93} 
eliminating a terrestrial source also rules out significant solar 
cosmic ray spallation production.  Thus, two independent 
measurements with high signal to noise (shown by the 100 to 1 
leading/trailing surface ratio), establish that {\it PP chain 
fusion near the solar surface is the sole source capable of 
providing the requisite flux} but say naught about the fusion 
process. 

\vskip 4pt
\noindent RESOLUTION OF THE NEUTRINO ANOMALY -- With most (all?) 
fusion near the surface, many core premises of the SSM, e.g., 
long term parametric stability, are invalid.  With fusion a 
temporal and spatial variable, changes in isotopic-chemical 
fractions, temperature and pressure must cause variable 
branching.  Observation indicates that they are within several 
multiples of SSM values.  

Solar cycle variation of temperature and 
composition\cite{Livingston91} permits us to explain neutrino 
measurements.  Once formed, three possibilities exist for $\rm 
^7Be$; 1) electron capture, 2) proton capture or 3) departure in 
the solar wind.  During solar maximum, UV, EUV and x-ray 
luminosity increase dramatically, indicative of a substantial 
increase in high energy electrons.  This enhances PP$II$ 
reactions increasing $\rm ^7Be$ neutrinos.  If PP$II$ goes to 
completion by electron capture (Fig.~1) and a small fraction is 
lost to the solar wind, fewer $\rm ^7Be$ nuclei are available for 
PP$III$ proton capture.  Further, the additional $\rm ^7Li$ 
produced competes with $\rm ^7Be$ for proton capture, assuring 
less $\rm ^8B$.  With electron capture favored $\sim \! 10^3$ 
over proton capture, we are discussing only the tail of the 
distribution.  However, if the high energy electron density 
increases, producing more $\rm ^7Li$, and the solar wind rises, 
then the flux of $\rm ^8B$ neutrinos must decrease appreciably.  
This theoretically dominates Homestake measurement and is 
substantially all that Kamiokande can detect.  Such an anti-
correlation with sunspot number was reported\cite{Rowley85}.  
This results in a $\rm ^7Be$ neutrino flux that follows the solar 
cycle, anti-correlated with $\rm ^8B$ neutrinos.  The neutrino 
energy dependence of $\rm ^{37}\hskip -1pt Cl$ capture heavily 
weights $\rm ^8B$ neutrino detections in Homestake data 
(Table~I), indicating why nearly simultaneous measurements must 
be used to compare Homestake with Kamiokande\cite{Davis94}.   

Measurement shows that PP fusion is enhanced during major flares 
and branching should vary significantly.  While Kamiokande data 
may vary with the solar cycle\cite{Davis94}, it is unclear 
whether real time Kamiokande $\rm ^8B$ neutrino detections 
increased during the 1991 activity.  Apart from absence of 
reports, our uncertainty stems from the reduced $\rm ^8B$ 
neutrino flux measured during solar maxima.  It is possible for a 
fusion burst, while multiplying the $\rm ^7Be$ neutrino flux, to 
leave unchanged or even reduce $\rm ^8B$ neutrinos. 

The missing $\rm ^7Be$ ``paradox'' results from assuming an 
invariant neutrino flux\cite{Bahcall92}.  LDEF and solar spectral 
observations attest to an adequate $\rm ^7Be$ supply.  In showing 
contemporaneous consistency with Kamiokande, Davis\cite{Davis94} 
reduced Homestake detections by 0.77 (Table~I) to compute the 
$\rm ^8B$ flux, leaving room for a $\rm ^7Be$ component.  The 
``missing $\rm ^7Be$'' component of recently reported SAGE and 
GALLEX values are due to currently reduced solar fusion and 
should get even lower when solar minimum data are added before 
beginning to again rise (if GALLEX continues). 

\vskip 4pt
\noindent SUMMARY AND DISCUSSION -- The five elements presently 
comprising the solar neutrino problem: 1) neutrino shortfall; 2) 
neutrino solar cycle variation; 3) the Homestake Kamiokande 
discrepancy; 4) neutrino surges coinciding with major solar 
flares; 5) absence of $\rm ^7Be$ neutrinos; measurements that 
diverge from prediction show the SSM flawed.  Each may be 
resolved with known physics recognizing the entire PP fusion 
chain occurs near the solar surface.   The shortfall relative to 
SSM prediction results from the core (fusion) assumption of long 
term invariance and its concomitant effect upon branching.  
Changes in $\rm ^3He$ (PP$II$) and energetic electron 
concentrations (PP$III$), observed on the active sun, cause 
variable branching resulting in the Homestake solar cycle 
anti-correlation.  The PP and $\rm ^7Be$ neutrino flux correlate 
with sunspot number showing varying fusion.  Enhanced PP$II$ 
terminations and solar wind loss reduce $\rm ^8B$ production, 
causing the PP$III$ neutrino flux to change inverse to solar 
activity.  While the $\rm ^8B$ neutrino flux is about three 
orders of magnitude less than from $\rm ^7Be$ (Table~I), 
detection by $\rm ^{37}\hskip -1pt Cl$ is near comparable due to 
the large neutrino energy dependence.  Hence, we conclude that 
the Homestake detections of $\rm ^7Be$ neutrinos relative to $\rm 
^8B$ detections vary with solar activity.  While the Homestake 
Kamiokande ``discrepancy'' has been resolved\cite{Davis94}, long 
term comparison of data should establish this difference in the 
two neutrino fluxes. 

Bursts of neutrino detections recorded during active solar 
periods are the consequence of multiplied near surface PP fusion 
during coronal mass ejections and associated major solar flares.  
These spikes are substantial in Homestake, and may have been 
detected by GALLEX and SAGE during the 1991 active period. 
Variation of Homestake data with sunspot number is significant 
and indicates cycling near surface fusion rates.  Changes in 
GALLEX and (allowing for start-up problems) in SAGE,  are 
consistent with solar cycle variation of PP surface fusion, 
contradicting the SSM. 
  
In summation, present measurements give no credence to fusion 
occurring in the core region of the sun and therefore, stars in 
general.  Commonly observed large and short period stellar 
luminosity variations also support fusion near the stellar 
surface. 

Noteworthy is that the foregoing is independent of neutrino mass 
and magnetic moment.  Other non-standard models generally address 
one aspect of the anomaly, e,g,. the deficient flux.  This near 
surface fusion solution is unique in respecting all measurements 
and in particular recognizing the significance of the LDEF $\rm 
^7Be-^{10}\! Be$ results.  A serious criticism must supply an 
alternative to fusion near the solar surface as the $\rm ^7Be$ 
source.  Beyond the limits of this communication is discussion of 
near surface (below excepting major flares), non 
equilibrium fusion mechanics, a subject of future papers.

\begin{figure}
\caption{Principle reactions, approximate Standard Solar Model 
branching (relative to 100\% proton-proton combinations) and 
designations for the solar hydrogen to helium fusion chain.}
\label{Fig. 1}
\end{figure}

\begin{figure}
\caption{Chronology of GALLEX and SAGE reported results.  
Early SAGE points are for the time periods indicated by 
horizontal bars, while all GALLEX points are means since 1991 
startup.  Errors shown are statistical and systematic added 
quadratically. At the upper right is shown the range of 
predictions of derivations from the Standard Solar Model (closed 
bar) with their $1\sigma$ error extensions (open bars).  The 
horizontal line is the theoretically computed minimal solar model 
below which there is insufficient hydrogen fusion to power the 
solar luminosity [10].  Early low values reported by the SAGE 
consortium, believed to result from start-up difficulties, were 
subsequently revised upward [7].}
\label{Fig. 2}
\end{figure}

\begin{table}
\caption{Standard Solar Model neutrino predictions [18] and 
measurements [3] compared.} 
\label{Table I}
\begin{tabular} {crlcdldldl}
$\nu$ & Branching & \hskip 3em $\nu$ energy & $\nu$ flux  
& \multicolumn {2} {c} {Homestake} & \multicolumn {2} {c} {SAGE} 
\& & \multicolumn {2} {c} {Kamiokande}\\ 
Source & \hskip -1em Ratio &&&&& \multicolumn {2} {c} {GALLEX} 
& \multicolumn {2} {c} {7.5 MeV cut}\\
Reaction &&&& \multicolumn {2} {c} {$\rm ^{37}\hskip -1pt 
Cl$\tablenotemark [1]} & \multicolumn {2} {c} {$\rm 
^{71}Ga$\tablenotemark [1]} & \multicolumn {2} {c} {$\rm 
^8B\,\phi$\tablenotemark [2]}\\ 
&& \hskip 3.5em (MeV) & ($\rm m^{-2}s^{-1}$) 
& \multicolumn {2} {c} {(SNU)} & \multicolumn {2} {c} {(SNU)} 
& \multicolumn {2} {c} {$\rm (10^{-2} m^{-2}s^{-1})$}\\ 
\tableline
PP & $\sim\! 100\%$ & 0 -- 0.420 spectrum & $6.0[10]^{14}$ 
&  0 && 71 && 0 &\\
$\rm ^7Be$ & 13\% & 0.861 line (90\%) & $4.7[10]^{13}$ 
& 1.1 && 34 && 0 &\\ 
&& 0.383 line (10\%) &&&&&&&\\
$\rm ^8B$ & 0.017\% & 0 -- 14.1 spectrum & $5.8[10]^{10}$  
& 6.1 && 14 && 5.7 &\\ 
\tableline
Total\tablenotemark [3] &&&& 7.9 && 132 && 5.7 &\\
\tableline
Measured &&&& 2.55 & $\pm 0.25$ & 75 & $\pm 7$ & 2.9 & $\pm 0.4$ \\ 
\end{tabular}
\tablenotetext [1] {Threshold for $\rm ^{37}\hskip -1pt 
Cl(\nu,\,e^-)^{37}\hskip -2pt Ar$ is 0.814 Me\hskip -2pt V and 
for $\rm ^{71}Ga(\nu,\,e^-)^{71}Ge$ is 0.233 Me\hskip -1pt V.} 
\tablenotetext [2] {As the SNU refers to captures, fluxes are 
listed for the Kamiokande electron scatter \v Cerenkov detector.}  
\tablenotetext [3] {As only principle solar hydrogen fusion 
reactions are listed, totals are larger than the sum of the 
values shown.} 
\end{table}


\begin{references}

\bibitem[*]{*} To whom correspondence should be addressed.  
    E-mail: rsoberma@mail.sas.upenn.edu. 

\bibitem{Davis68} R. Davis Jr., D. S. Harmer, and K. C. Hoffman,   
    Phys. Rev. Lett. {\bf 20}, 1205 (1968).

\bibitem{Bahcall89} J. N. Bahcall, {\it Neutrino Astrophysics} 
    (Cambridge U., Cambridge, 1989). 

\bibitem{Bahcall95} J. N. Bahcall {\ it et al.}, Nature {\bf 
    375}, 29 (1995). 
      
\bibitem{Hirata91} K. S. Hirata {\it et al.}, Phys. Rev. D {\bf 44}, 
    2241 (1991). 
      
\bibitem{Abazov91} A. I. Abazov {\it et al.}, Phys. Rev. Lett. {\bf 67}, 
    3332 (1991). 
      
\bibitem{Anselmann92} P. Anselmann {\it et al.}, Phys. Lett. B
    {\bf 285}, 376 (1992).
   
\bibitem{Abdurashitov94} J. N. Abdurashitov {\it et al.},  
    Phys. Lett. B {\bf 328}, 234 (1994). 
  
\bibitem{Anselmann93} P. Anselmann {\it et al.}, Phys. Lett. B
    {\bf 314}, 445 (1993).
  
\bibitem{Anselmann95} P. Anselmann {\it et al.}, Phys. Lett. B 
    {\bf 357}, 237 (1995).

\bibitem{Bahcall85} J. N. Bahcall, B. T. Cleveland, R. Davis Jr., and 
    J. K. Rowley, Astrophys. J. Lett. {\bf 292}, L79 (1985). 
  
\bibitem{Rowley85} J. K. Rowley, B. T. Cleveland, and R. Davis Jr., 
    in {\it Solar Neutrinos and Neutrino Astronomy (Homestake, 1984)} 
    {\bf Conf. Proceed. No. 126}, p. 1 (A. I. P., New York, 1985).   

\bibitem{Bahcall87} J. N. Bahcall, G. B. Field, and W. H. Press,   
    Astrophys. J. Lett. {\bf 320}, L69 (1987).  

\bibitem{Davis94} R. Davis Jr., Prog. Part. Nucl. Phys. {\bf 32}, 13 
    (1994).

\bibitem{Baziliv82} G. A. Bazilivskaya, Yu. I. Stozhkov, and 
    T. N. Charakhch'yan, JETP Lett. {\bf 35}, 341 (1982); 
    R. Davis Jr., in {\it Proc. Seventh Workshop on Grand
    Unification, ICOBAN'86 Toyama, Japan}, edited by J. Arafune, 
    p. 237 (World Scientific, Singapore, 1987).

\bibitem{Gurnett93} D. A. Gurnett, W. S. Kurth, S. C. Allendorf, 
    and R. L. Poynter, Science {\bf 262}, 199 (1993).

\bibitem{Webber93} W. R. Webber  and J. A. Lockwood, J. Geophys. 
    Res. {\bf 98}, 7821 (1993).

\bibitem{Bahcall94} J. N. Bahcall, Phys. Lett. B {\bf 338}, 276 (1994);
    R. S. Raghavan, Science {\bf 267}, 45 (1995).
      
\bibitem{Bahcall92} J. N. Bahcall and M. H. Pinsonneault, Revs. 
    Mod. Phys. {\bf 64}, 885 (1992).

\bibitem{MSW} S. P. Mikheyev and A. Yu. Smirnov, Nuovo 
    Cimento {\bf 9C}, 17 (1986).

\bibitem{Sienk90} R. Sienkiewicz, J. N. Bahcall, and B. Paczynski,  
    Astrophys. J. {\bf 349}, 641 (1990).

\bibitem{Fishman91} G. J. Fishman {\it et al.}, Nature {\bf 349}, 
    678 (1991); G. W. Phillips {\it et al.}, in {\it LDEF - 69 
    Months in Space, Proceedings of the First Post-Retrieval 
    Symposium} {\bf CP-3134}, p. 225 (NASA, Washington, 1992).                  

\bibitem{Arnold55} J. R. Arnold and H. A. Al-Salih, Science {\bf 121}, 
    451 (1955).
      
\bibitem{Gregory93} J. C. Gregory {\it et al.}, in {\it LDEF - 69 Months in 
    Space, Proceedings of the Second Post-Retrieval Symposium} 
    {\bf CP-3194}, p. 231 (NASA, Washington, 1993).

\bibitem{Chappell87} C. R. Chappell, T. E. Moore, and J. H. Waite Jr., 
    J. Geophys. Res. {\bf 92}, 5896 (1987). 

\bibitem{Terekhov93} O. V. Terekhov {\it et al.}, Astron. Lett.  
    {\bf 19}, 65 (1993). 

\bibitem{Livingston91} W. Livingston {\it et al.}, in {\it Solar 
    Interior and Atmosphere}, edited by A. N. Cox, 
    W. C. Livingston, and M. S. Matthews, p. 1109 (U. of Ariz., 
    Tucson, 1991).  

\end{references}
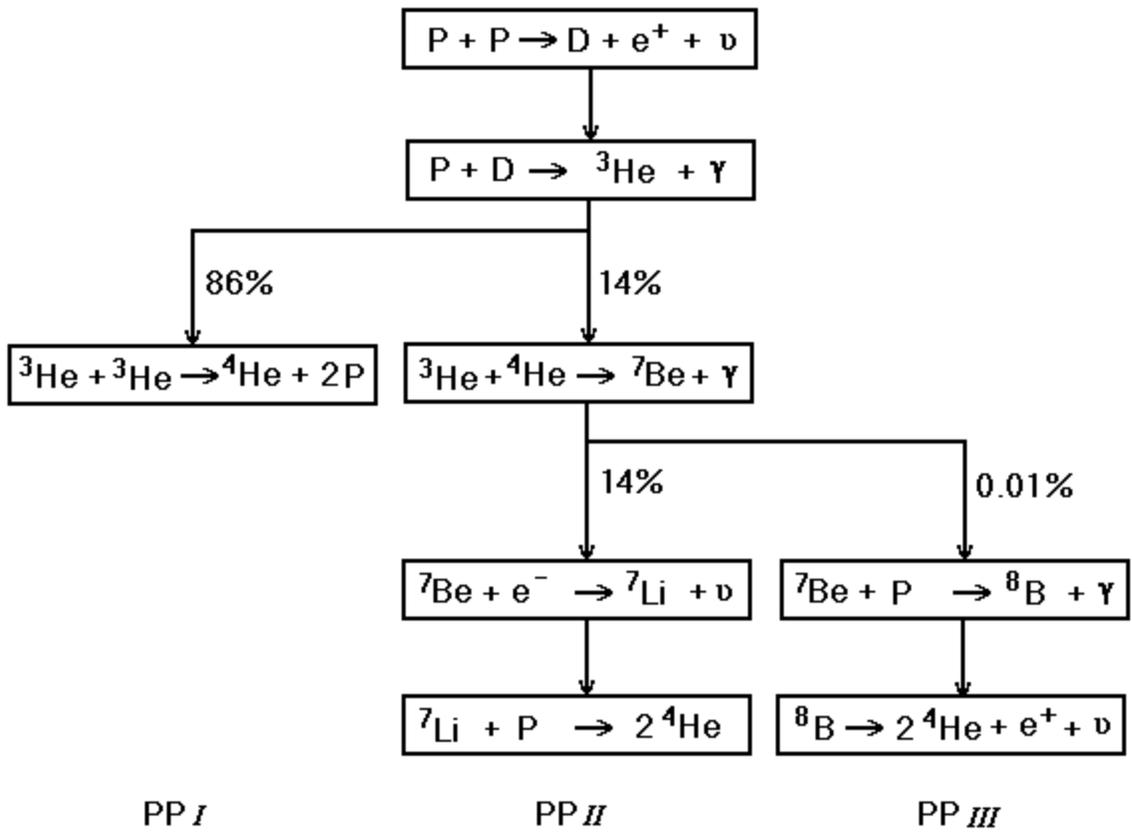
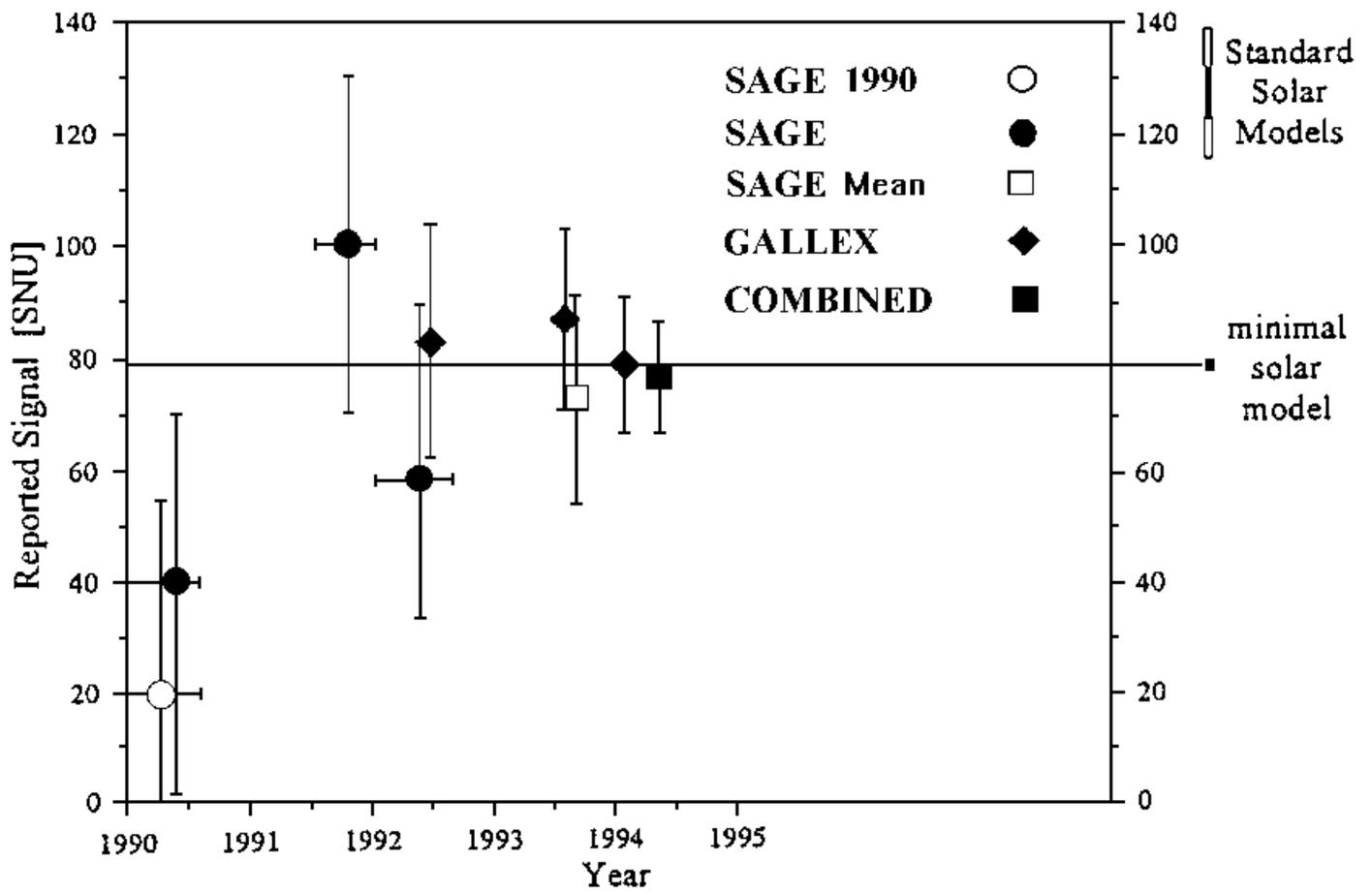
\end{document}